\input harvmac
\input epsf

\def\R{\relax{\rm I\kern-.18em R}}
\font\cmss=cmss10 \font\cmsss=cmss10 at 7pt
\def\Z{\relax\ifmmode\mathchoice
{\hbox{\cmss Z\kern-.4em Z}}{\hbox{\cmss Z\kern-.4em Z}}
{\lower.9pt\hbox{\cmsss Z\kern-.4em Z}}
{\lower1.2pt\hbox{\cmsss Z\kern-.4em Z}}\else{\cmss Z\kern-.4em
Z}\fi}\
\def\np{Nucl. Phys. }
\def\pl{Phys. Lett. }
\def\pr{Phys. Rev. }
\def\lmp{Lett. Math. Phys. }

\def\cmp{Comm. Math. Phys. }
\def\mpl{Mod. Phys. Lett. }
\def\p{\partial}
\def\CC{{\cal C}}
\def\CF{{\cal F}}

\def\CZ{{\cal Z}}

\def\l{\ell}

\def\CD{{\cal D}}

\def\MG{{\cal M}_G}
\def\CM{M_G}
\def\Tr{{\rm Tr}}

\def\f{\Phi}
\def\char{\chi_{_h}}

\def\sumh{\sum_{h}}
\def\CN{{\cal N}}

\def\Oh{\Omega_{h}}
\def\ov{\over}
\def\dag{\scriptscriptstyle \dagger}
\def\barint{-\hskip -11pt\int}
\def\half{{1 \over 2}}
\def\CN0{{\cal N}_0}
\def\t{\tau}

\rightline{CERN-TH/97-53}
\rightline{hep-th/9703189}

\Title{}
{\vbox{
\centerline{Complex Matrix Models}
\centerline{and} 
\centerline{Statistics of Branched Coverings of 2D Surfaces}
}}

\vskip6pt

\centerline{Ivan K. Kostov$^{\ast}$\footnote{$^\diamond $}{member of CNRS}\footnote{$ ^\dagger$}{{\tt kostov@spht.saclay.cea.fr}} , Matthias Staudacher$^\natural$\footnote{$^\bullet$}{{\tt
matthias@nxth04.cern.ch}} and   Thomas Wynter$^{\ast}$\footnote{$^\circ$}{{\tt wynter@spht.saclay.cea.fr}}
  }

\centerline{*{\it  Service de Physique Th\'eorique, C.E.A. - Saclay, F-91191 Gif-Sur-Yvette, France}}

\smallskip

\centerline{{$\natural$\it CERN, Theory Division, CH-1211 Geneva 23, Switzerland}}

\vskip .3in
\baselineskip10pt{
We present a complex matrix gauge model defined on an arbitrary
two-dimensional orientable lattice. We rewrite the model's partition
function in terms of a sum over representations of the group
$U(N)$. The model solves 
the general combinatorial problem of counting
branched covers of orientable Riemann surfaces with any given,
fixed branch point structure. We then define an appropriate 
continuum limit allowing the branch points to freely float 
over the surface. The simplest such limit reproduces
two-dimensional chiral $U(N)$ Yang-Mills theory and its string description
due to Gross and Taylor.
}
\bigskip

\rightline{SPhT-97/022}
\leftline{CERN-TH/97-53}
\Date{March 1997 }

\baselineskip=16pt plus 2pt minus 2pt
\bigskip
\lref\CWM{I.~Kostov and M.~Staudacher, hep-th/9611011,
to be published in Phys. Lett. B}
\lref\WG{D.~Weingarten, \pl B 90 (1980) 280.}
\lref\EGU{T.~Eguchi and H.~Kawai, \pl B 114 (1982) 247
and \pl B 110 (1982) 143. }
\lref\DFJ{B.~Durhuus, J.~Fr\"ohlich and T.~J\'onsson,
\np B 240 FS[12] (84) 453.}
\lref\KSW{V.A.~Kazakov, M.~Staudacher and T.~Wynter,
\cmp 177 (1996) 451; \cmp 179 (1996) 235;
\np B 471 (1996) 309.}
\lref\AFOQ{H. Awada, M. Fukuma, S. Odake and Y.-H. Quano, \lmp 31
(1994) 289.}
\lref\KSWII{V.A.~Kazakov, M.~Staudacher and T.~Wynter,
\cmp 179 (1996) 235.}
\lref\KSWIII{V.A.~Kazakov, M.~Staudacher and T.~Wynter,
\np B 471 (1996) 309.}
\lref\KSWR{V.A.~Kazakov, M.~Staudacher and T.~Wynter,
{\it Advances in Large $N$ Group Theory
and the Solution of Two-Dimensional $R^2$ Gravity},
hep-th/9601153,
1995 Carg\`ese Proceedings.}
\lref\GAUGE{A.A. Migdal, Zh. Eksp. Teor. Fiz. 69 (1975) 810
(Sov. Phys. JETP 42 (413)).}
\lref\DK{M.R.~Douglas and V.A.~Kazakov, \pl B 312 (1993) 219.}
\lref\RUS{B.~Rusakov, \mpl A5 (1990) 693.}
\lref\GT{D.~Gross, \np B 400 (1993) 161;
D.~Gross and W.~Taylor, \np B 400 (1993) 181;
\np B 403 (1993) 395.}
\lref\MOO{S.~Cordes, G.~Moore and S.~Ramgoolam,
{\it Large N 2-D Yang-Mills Theory and Topological String Theory},
hep-th/9402107, and
{\it Lectures on 2D Yang-Mills Theory,
Equivariant Cohomology and Topological Field Theories},
hep-th/9411210, 1993 Les Houches and Trieste Proceedings;
G.~Moore, {\it 2-D Yang-Mills Theory and Topological Field Theory},
hep-th/9409044.}
\lref\Hv{P.~Ho\v rava, \np B 463 (1996) 238.}
\lref\GW{D.J.~Gross and E.~Witten, \pr D 21 (1980) 446.}
\lref\IKK{V. Kazakov, \pl B 128 (1983); K. O'Brien and J.-B. Zuber,
\pl B 144 (1984) 407; I.~Kostov, \np B 265 (1986), 223, B 415 (1994)
29.}
\lref\DOUG{M.R.~Douglas,
{\it Conformal Field Theory Techniques in
Large $N$ Yang-Mills Theory}, hep-th/9311130, 1993 Carg\`ese
Proceedings.}
\lref\RUDD{R.~Rudd, {\it The String Partition Function for
QCD on the Torus}, hep-th/9407176.}
\lref\DIJKI{R.~Dijkgraaf, {\it Mirror Symmetry and Elliptic Curves},
in
{\it The Moduli Space of Curves}, Progress in Mathematics 129
(Birkh\"auser, 1995), 149.}
\lref\DIJKII{R.~Dijkgraaf, {\it Chiral Deformations of Conformal
Field Theories}, hep-th/9609022.}
\lref\ITDI{P.~Di~Francesco and C.~Itzykson, Ann. Inst. Henri.
Poincar\'e Vol. 59, no. 2 (1993) 117.}
\lref\K{V.A.~Kazakov, \np B 354 (1991) 614.}
\lref\KW{V.A.~Kazakov and T.~Wynter, \np B440 (1995) 407.}
\lref\KSWIV{I.~Kostov, M.~Staudacher and T.~Wynter,
in preparation.}
\lref\KK{V.A.~Kazakov and I.~Kostov,
\np B 176 (1980) 199; V.A.~Kazakov, \np B 179 (1981) 283.}
\lref\WC{W.~Taylor and M.~Crescimanno \np B 437 (1995) 3.}
\lref\Wati{W.~Taylor, MIT-CTP-2297 hep-th/9404175 (1994).}
\lref\GW{D.~Gross and E.~Witten, Phys.Rev. D21 (1980) 446.}
\lref\GB{D.~Gross and E.~Br\'ezin, \pl B 97 (1980) 120.}
\lref\MW{M.~Staudacher and T.~Wynter, unpublished (1996).}
\lref\Kaz{V.~Kazakov, \pl B 128 (1983) 316.} 

\newsec{Introduction}
  
Recently, two of the authors have considered a complex 
matrix model which describes the ensemble of branched coverings  of a two-dimensional manifold \CWM.  The model has been interpreted as a string theory  invariant with respect to area-preserving diffeomorphisms of the target space. 
In this paper, we continue the investigation of this model.

The geometrical problem we solve consists in the enumeration of the branched coverings of a
two-dimensional manifold with given number of punctures. The target manifold is characterized by its topology, the number of punctures, and its total area.   All structures we are considering are invariant under area-preserving diffeomorphisms of the target manifold.
We assume that the covering surfaces can have branch points located at the punctures. Two covering surfaces related by an area-preserving diffeomorphisn are considered identical.  

The problem will be reformulated in terms of a lattice gauge theory of
$N\times N$ complex matrices defined on a lattice representing 
 a cell decomposition of the target manifold.  The vertices  of the lattice are the punctures of the target surface. 
 The ${1 \over N}$ perturbative expansion of this model generates the covering surfaces with the corresponding combinatorial factors.  These discrete surfaces can be interpreted as cell decompositions of continuum surfaces covering the target space. It should be intuitively clear that the combinatorics of these surfaces
  should not depend on the cellular decomposition
 of the target manifold, but only on global features like
the topology of the manifold, the topology of the covering
Riemann surface, and the number and types of branch points
allowed. Our solution below will confirm this intuition.

The solution of the model is given in terms of a sum over
polynomial representations of the group $U(N)$. This sum is,
by construction, a generating function for the combinatorics
of covering maps. The order
of the representation gives the degree of the (connected and
disconnected) coverings. The sum depends on the genus $G$ of the target
manifold, and a number of variables tied to geometrical data
of the covering maps: $N^{-2}$ is the  genus
expansion parameter of the covering surfaces, and we associate, for each cell corner $p$, a 
weight $t_k^{(p)}$ corresponding to a branch point at $p$ of order $k$.

For some applications, one would consider a slightly more general 
problem of counting  coverings  with branch points that can occur  
anywhere on the smooth target manifold. In order to allow for this
possibility, we are led to take a {\it continuum limit}:
We simply cover the target manifold by a microscopically small
cell decomposition, with the weights of these branch points tuned correspondingly.   We will find that the combinatorics of the 
resulting statistics of movable branch points actually simplifies
significantly over the general case of fixed branch points:
Many special configurations of enhanced symmetry, corresponding
to coalescing branch points, are scaled away.  
The weights associated with the punctures  are not
subdued to  scaling; this makes the difference between the punctures and the rest of the points of the  cell decomposition.
   
Apart from the obvious mathematical interest of our approach,
we are able to connect our results to recent work on the 
QCD string in two dimensions. In the case of a target space with nonnegative global curvature, the chiral (i.e.,
orientation preserving) sector of the Gross and Taylor \GT\ string theory describing two dimensional Yang-Mills theory,
is a particular case of the string theory   defined by our matrix model.   We discuss the problem of the 
``$\Omega$ factors'' in the string interpretation of the Yang-Mills theory and interpret   the "$\Omega$-points" as punctures in the target space.

\newsec{Definition of the model}

Consider a smooth, two dimensional, compact, closed and
orientable manifold $\MG$  of genus $G$  with ${\cal N}_0$  marked points (punctures), which we denote by  $p=1,...,  {\cal N}_0  $. The manifold is compact in the following sense: We assume that there is a volume form $dA$ on $\MG$
and the total area $A_T=\int dA$ is finite.  We will consider the ensemble of nonfolding surfaces
covering $\MG$ and allowed to have branch points at the punctures.  These surfaces are
smooth everywhere on $\MG$ and are given a volume form inherited from the embedding. In this way, the area of a surface covering $n$ times the target manifold is equal to $nA_T$.
 
 We will resolve the problem by discretizing  the target manifold.  Introduce a cell decomposition  of the original target manifild  $\MG$   such that  each cell
is a polyhedron homeomorphic to a disc. The vertices of the cell decomposition are by construction the $\CN0$ punctures of $\MG$.   
The resulting polyhedral surface (cellular complex)  $\CM$ is thus 
characterized by its genus $G$, and by its set of  points
$p$, links $\ell$ and cells $c$. The numbers of points, links and cells which we denote correspondingly by ${\cal N}_0$, ${\cal N}_1$ and
${\cal N}_2$, are related by the Euler formula
\eqn\eiL{ 
{\cal N}_0-{\cal N}_{1} +{\cal N}_2 = 2-2G.
}
Each cell contributes a fraction $A_c$ to the total area $A_T$
of the target manifold, so that $A_T=\sum_c A_c$.
 A given manifold can be discretized in many different ways, but the choice of discretization is irrelevant for our problem.  
 
A branched covering of $\CM$ of degree $n$
is a surface $\Sigma$ obtained by taking $n$ copies of each
of the polygons of $\CM$ and identifying pairwise the edges of the
$n$ polygons on either side of each link.  The Riemann surface
obtained in this way can have branch points of order $k$ 
($k=1,2,...,n$) representing cyclic contractions of edges. They are
located at the points  $p \in \CM$.  The discretized surface $\Sigma$
has $n {\cal N}_1$ links, $n {\cal N}_2$ polygons and $n {\cal N}_0-\sum_p b_p$ points (with
$b_p$ the winding number minus one at point $p$). Its
total area is $n A_T$. Its genus $g$ is
given by the Riemann-Hurwitz formula:
\eqn\riemann{
2 g-2= n(2 G-2)+\sum_p b_p.}
The partition function is defined as the sum over all possible  coverings $\Sigma\to \CM$
conserving the orientation. A factor $N^{2-2g}$ is assigned to the genus $g$  of the covering surface. Furthermore, we introduce Boltzmann weights  associated with its branch points. The weight
of a branch point of order $k$ is $t_k^{(p)}$, where $k\geq2$.
A regular (analytic) point gets a weight $t_1^{(p)}$.

The partition sum is now defined as a sum over all
(not necessarily connected) coverings $\Sigma\to \CM$:
\eqn\latZpr{
Z=\sum_{\Sigma\to \CM} e^{-n A_T} N^{2-2g}
\prod_{p=1}^{{\cal N}_0} \prod_{k\ge 1} (t_k^{(p)})^{n_k}.}
where, associated with the point $p\in \CM$,
$n_k(p)$ (with $k\geq 2$) is the number of the branch points of order $k$ of $\Sigma$ and $n_1(p)$ the number of regular points.
The symmetry factor  of the map is understood in the sum.

Now we introduce a matrix model whose perturbative expansion
coincides with \latZpr.
To each link $\l=\langle pp' \rangle$ we associate a field variable $\Phi_\l$
representing an $N\times N$ matrix with complex elements. By
definition $\Phi_{<pp'>}= \Phi_{<p'p>}^{\dag} $.  In order to be able
to associate arbitrary weights to the branch points we will
associate an external matrix field with the corners of the cells. Let
us denote by $(c, p>$ the corner of the cell $c$ associated with the
point $p$.  The corresponding matrix will be denoted by $B_{(c,p>}$.

\vskip 50pt
\hskip 20pt
\epsfbox{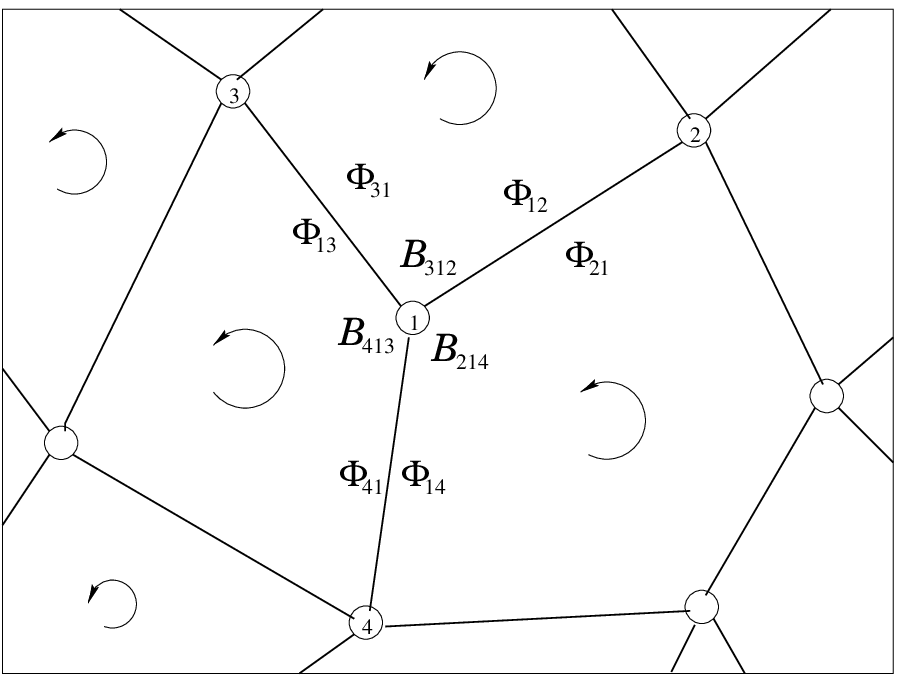}
\vskip 20pt
\centerline{Fig.1: The matrices associated with the links 
$<12> ,<13> ,<14>$ }
\centerline{and the corners $(214>,(312>, (413>$ associated with  the point 1.   }

\bigskip

The  partition function of the matrix model is defined as
\eqn\cwm
{
Z=\int \prod_{\l} [\CD \f_{\l}]~\prod_c
{}~\exp (e^{-A_c}~N~\Tr   \f_{c} ), 
}
where  $\Phi_c$ denotes the ordered product of link and 
corner variables along the oriented boundary $\p c$ of the  
cell $c$
\eqn\phhic{
\Phi_c=\prod_{\l , p\in \p c}\Phi_\l B_{(c,p>},}  
 and the integration over the link variables is performed with the
Gaussian measure
\eqn\measure{
 [\CD \Phi_{\l}]=({N /\pi})^{N^2}
\prod_{i,j=1}^{N} d(\Phi_{\l})_{ij} d(\Phi_{\l})^{*}_{ij}
~e^{-N~\Tr~\Phi_{\l}^{ } \f_{\l}^{\dag} }.
}

The perturbative expansion of \cwm\ gives exactly the partition
function \latZpr\ of branched surfaces covering $\CM$. 
The weight $t_k^{(p)}$
of a branch point of order $k \ (k\ge 2)$ or regular point ($k=1$)
associated with the vertex
$p\in\CM$ equals
\eqn\cpL{
t_k^{(p)}={1\over N}  \Tr [B_p^k],}
where we have defined the matrices $B_p$ as the
ordered product
\eqn\prBB{
B_p = \prod_{c}B_{(c,p>}
}
of the $B$-matrices around the  vertex $p$. 

%
%

\newsec{Exact solution by the character expansion method}

The method consists in replacing the integration over complex 
matrices by a sum over polynomial representations of $U(N)$.
Applying to \cwm\ the same strategy as in ref.\KSWR , we expand the
exponential of the action for each cell $c$ 
as a sum over the Weyl characters $\char$ of these representations:
\eqn\charexp{
\exp (e^{-A_c}~N~\Tr \f_{c} ) = \sum_{ h } {\Delta_h \ov \Omega_h}~\char(\f_c)~
e^{-A_c |h|}.} 
The representations are parametrized by the shifted weights
$h=\{h_1,h_2,\ldots,h_N\}$, where $h_i$ are related to the lengths
$m_1,..., m_N$ of the rows of the Young tableau by $h_i=N-i+m_i$ and
are therefore subjected to the constraint $h_1>h_2>\ldots>h_N\geq 0$. We will  denote by $|h| = \Sigma_i m_i$ the total number of boxes of
the Young tableau. 
The dimension $\Delta_h$ of the representation $h$ is given by
\eqn\dim{\Delta_h=\prod_{i<j} { h_i-h_j \ov  j-i},}
and the ``Omega factor'' $\Omega_h$ by
\eqn\omeGa{\Omega_h  = N^{-|h|}~\prod_{i=1}^{N}
{h_{i}! \ov (N-i)!}.}
An explicit representation
of the Weyl characters $\char$ is only needed for the derivation
of two essential integration formulas (fission and fusion
rule, respectively):
\eqn\cff{\eqalign{
\int [\CD \f]~\char(\f_1~\f~\f_2~\f^{\dag})&=
{\Omega_h \over \Delta_h}
{}~\char(\f_1)~\char(\f_2), \cr
\int [\CD \f]~\char(\f_1~\f)~\chi_{_ {h'}}(\f^{\dag}~\f_2) & =
\delta_{h, h'}~{\Omega_h \over \Delta_h}~\char(\f_1~\f_2). \cr }}
For a simple proof of these identities, see \CWM.
It is now possible to exactly perform the integration with respect to the 
gaussian measure \measure\ at each link. 
Using the fusion rule one progressively eliminates links
between adjoining cells until one is left with
a single  cell.
The remaining links along that plaquette are integrated
out by applying the fission rule. A similar procedure
was first used in the exact solution of two-dimensional
Yang-Mills theory \GAUGE,\RUS. Employing the very useful relation 
\eqn\aone{
\char(A_1)={\Delta_h \ov \Omega_h} \;\;\;\;{\rm with}\;\;\;\;
{1 \over N}~\Tr A_1^k=\delta_{k,1}~,}
one finds the
exact solution of the matrix model \cwm:
\eqn\solution{
Z=\sumh 
\ \bigg({\Delta_h \over \Oh}\bigg)^{2-2 G}~     
\prod_{p=1}^{\CN0}~\bigg[  {\char(B_p ) \over \char(A_1)}
\bigg]~e^{-|h| A_T}. }
The characters $\char(B_p)$ are related to the branch point
weights $t_k^{(p)}$ at the site $p$ through the Frobenius formula
(we omit the index $p$ for clarity):
\eqn\frobenius{
\char(B)=\sum_{n_1+2 n_2+3 n_3 \ldots=|h|} 
{\rm ch}_h(1^{n_1},2^{n_2},3^{n_3},\ldots)~
{|h|!~N^{n_1+n_2+n_3+\ldots}
 \over 1^{n_1} n_1!~2^{n_2} n_2!~3^{n_3} n_3! \ldots}~
t_1^{n_1} t_2^{n_2} t_3^{n_3}\ldots}
The sum is over all partitions of the covering number
$|h|$, and the ${\rm ch}_h(1^{n_1},2^{n_2},3^{n_3},\ldots)$ 
are the characters of the symmetric group $S_{|h|}$ corresponding
to the representation $h$ and the partition class with
cycle structure $(1^{n_1},2^{n_2},3^{n_3},\ldots)$ . 
Let us emphasize that \solution, together with \frobenius,
is the complete solution of the combinatorial problem
of counting branched covers \latZpr.

In appendix A, we have given the first few terms of the 
expansion \solution, in the special case where the 
branch point weights are identical on all $\CN0$ sites.
We have also given the first few terms of the free energy $F=\log Z$,
corresponding to connected surfaces. 

A special case of \solution\ is obtained when 
 no branch 
points are allowed on the target manifold. 
This is done by
choosing all matrices $B_p=A_1$. Then the solution \solution\
becomes completely independent of the cell decomposition
of the manifold: The cell corners are indistinguishable from
any other point on the surface. We obtain
\eqn\nobranch{
Z=\sumh 
\ \bigg({\Oh \over\Delta_h}\bigg)^{2 G-2}~     
e^{-|h| A_T}. }
This is completely trivial for the sphere and very simple 
(see \CWM,\GT) for the torus, but highly non-trivial for higher
target space genus $G \geq 2$:
\eqn\topol{
\log Z=\cases{N^2~e^{-A_T}, &  for  $G=0$;\cr
N^0~\sum_{k=1,m=1}^{\infty}~{1 \over k}~e^{-k m A_T}  & for $G=1$;\cr
N^{2-2 G}~e^{-A_T} + (N^{2-2 G})^2~\half (4^G -1) e^{-2 A_T}
+  \ldots & for $G\geq 2$. \cr }}
For $G\ge 2$, $\log Z$ counts the number of smooth and locally invertible  maps  of a surface of genus $g$ 
onto a surface of genus $G$.  It would be interesting to find an analytic expression for
$\log Z$ for all $G\geq 2$.

\newsec{Continuum limit: statistics of movable branch points}

In the preceding section we have presented the full solution
to the problem of counting the branched covers of a smooth, 
closed and orientable target manifold of any genus $G$ and $\CN0$ punctures. 
By construction, the solution \solution\ depended, aside from
$G$ and the genus $g$ of the branched cover, on the number
and types of branch points at the punctures of $\MG$:  At each  puncture $p$
we are to specify a set of weights $t_k^{(p)}$ associated with the winding numbers of the covering surface. 
It is natural to consider a related, but different combinatorial problem:
 Allow the covering surfaces to have branch points  {\it anywhere} on the target manifold. 
In this case we have to sum over the positions of the branch points with an appropriate integral measure. 

The solution of this problem  is actually contained in the solution of the previous one.
Since branching
in our model is constrained to occur at the sites of the cell decomposition,
we are led to consider a continuum limit: The cell decomposition has
to densely cover the manifold.    The simplest choice is to  assign identical weights
 at each branching sites, that is, choose
$B_p=B$ everywhere, and assume that all  cells have the same area $A_c={A_T \over {\cal N}_0}$. 
Set 
\eqn\tntaum{\eqalign{ 
t_1=&{1 \over N} \Tr B=1 , \cr
t_k=&{1 \over N} \Tr B^k={\tau_k \over {\cal N}_0 }\ \ \ 
{\rm for} \ \ \ k\geq 2,}}
and take
${\cal N}_0\to\infty$, while holding $A_T$ and the
continuum couplings $\t_k$ fixed.
Thus the probablity for branching at a specific
site $p$ goes to zero in a prescribed way. 
In this continuum limit the configuration space of the 
branch points becomes 
infinite and their statistics drastically simplifies due to the fact
that  the probability to have more than one branch point  
associated with the same point of $\CM$ tends to zero.
Mathematically, this simplification is immediately seen
from the partition function \solution, and the Frobenius
formula \frobenius. The product of quotients of characters exponentiates,
and \solution\ becomes:
\eqn\solgen
{\CZ =\sum_{h}   
\bigg({\Delta_h\over\Omega_{h}}\bigg)^{2-2G} 
\exp\bigg[ \sum_{k=2}^{\infty} \tau_k~N^{1-k}\xi^h_{k}\bigg]~
e^{-|h| A_T}.}
Here, in view of \frobenius, the tableau dependent numbers
$\xi^h_k$ are given by
\eqn\xisym{
\xi^h_{k} ={|h|! \over k (|h| - k)!}
{{\rm ch}_h(1^{|h|-k},k^1) \over {\rm ch}_h(1^{|h|})}.}
An explicit formula is
\eqn\xidef{
\xi^h_{k}= 
{1 \over k } \sum_{i=1}^N h_i(h_i-1) \ldots (h_i-k+1) 
\prod_{{j=1 \atop j\neq i}}^{N}
\bigg(1- {k \over h_i - h_j}\bigg),}
which is quickly derived from \aone,\tntaum, and the identity
(coming from the Schur definition of the characters,
see \KSW)
\eqn\chids{
{\partial\over\partial t_k} \log \chi_h(B)=
{N\over k} \sum_{i=1}^N
{ \chi_{{\tilde {_h}}^{k}}(B) \over \char(B)}
\ \ \ \ {\rm with} \ \ \ \ {\tilde h}^{^k}_{i}=h_i-\delta_{k i}k.}
The $\xi_k^h$ are actually symmetric polynomials of degree $k$ in the
weights $h_i$ (see appendix B, where we have also listed a
few examples).
They are $N$ independent as long as $|h|<N$.

In appendix B, we have given the first few terms of the expansion
\solgen, as well as the first few terms of the free energy
$\CF=\log \CZ$, corresponding to connected surfaces.

%
%
%
%

Finally, let us mention that we can keep the weights associated with  
a given number of points unscaled.   Then we obtain the partition function of the ensemble of branched coverings of a target manifold  of area $A_T$, genus $G$ and $\CN0$ punctures. The  evident generalization of
eqs. \solution\ and \solgen\ is

\eqn\solutiongen{
Z=\sumh 
\ \bigg({\Delta_h \over \Oh}\bigg)^{2-2 G}~     
\prod_{p=1}^{\CN0}~\bigg[  {\char(B_p ) \over \char(A_1)}
\bigg]~\exp\bigg[ \sum_{k=2}^{\infty} \tau_k~N^{1-k}\xi^h_{k}\bigg]~e^{-|h| A_T}. }

\newsec{Relation to chiral $2D$ Yang-Mills theory}

The solution of the map counting problem with movable
branch points considered in the last section has an interesting
physical interpretation. Let us restrict ourselves to the case
of only simple branch points: $\t_k=0$ for $k\geq3$. Then, using
the explicit result for $\xi^h_2$ (see appendix B), and 
choosing
\eqn\tautwo{
\half \t_2=-~A_T,}
we can rewrite the partition function \solgen\ as
\eqn\ym{
\CZ =\sum_{h}   
\bigg({\Delta_h\over\Omega_{h}}\bigg)^{2-2G} 
~e^{- {A_T \over N}~C_2^h}.}
where $C_2^h$ is the second Casimir of the group $U(N)$.
This is very nearly the partition function of two-dimensional
$U(N)$ Yang-Mills theory. This theory was shown to be 
solvable on any target manifold in \GAUGE,\RUS. Recently,
it was demonstrated by Gross and Taylor \GT, that $YM_2$ can be 
interpreted as a string theory. This was achieved by the
{\it tour de force} approach of expanding the exact solution
in ${1 \over N}$ and interpreting the terms as string maps.
Here we are inverting the Gross-Taylor program: We start
with a theory whose interpretation as a string theory
generating covering maps from a worldsheet to the target
manifold is manifest, and aim to derive $YM_2$.
There remain, however, two subtle differences between \ym\
and $YM_2$:

(1) In $U(N)$ $YM_2$ the sum over polynomial representations $h$
is extended to a sum over coupled (non-polynomial) representations.
This difference is easy to understand: The missing representations
clearly correspond to orientation reversing maps, which have been
eliminated from the start by our chiral matrix model. Indeed it
became evident from the work of \GT, that $YM_2$ factorizes
into two rather weakly interacting chiral sectors. Our class of
models furnishes a precise realization of one such sector.  

(2) Our partition function contains, for genus $G\neq 1$,
the extra factor $\Omega_h^{2 G-2}$, absent in $YM_2$.
This factor {\it eliminates} the so-called ``$\Omega$ points''
and ``$\Omega^{-1}$ points''
introduced originally by Gross and Taylor in order to
sustain the string picture for $G\neq1$. 
For $G=0$, we can add two punctures and associate weight 1 with the windings 
around them (i.e. take a matrix source $B=1$). The answer is given by  \solutiongen\
with $\CN0=2$ and $B_1=B_2=1$ and coincide with the partition function of 
the Yang-Mills theory on the sphere.
 The two punctures  are the two 
``$\Omega$ points''. 
In summary, {\it in the case of a target space representing a sphere with two punctures
our theory is exactly equivalent\foot{ Adding a puncture to   the target space 
of the Yang-Mills theory does not change anything. 
For example, the gauge theory defined on a cylinder with Dirichlet boundaries is identical to the gauge theory on the sphere.}
to chiral $YM_2$.} 
 For $G\geq 2$ however, it is not possible
to eliminate the extra factor ( the ``$\Omega^{-1}$ points'')
by adding punctures. 
On the other hand, the ``$\Omega^{-1}$ factors'' have been given an interesting
re-interpretation in the work of Cordes, Moore and 
Ramgoolam \MOO. These authors showed that the $YM_2$
partition function could be rewritten as a sum
over covering maps {\it without} non-movable,
special singularities, but weighted instead with special
topological invariants (the ``Euler character of Hurwitz
moduli space''). 
 It would be interesting if this
result could be reproduced by a simple
matrix model, similar in spirit to \cwm\ and possessing an equally
clear surface interpretation.

\newsec{Sphere-to-sphere maps: saddle point analysis}

In the case of a target space of spherical topology $G=0$ we can 
apply a saddle point technique to the sum over representations to
extract the contribution from coverings with spherical topology (i.e. $g=0$).
This was explained in detail in \KSWR,\KSW, and
leads in the case of the model with fixed branch points \solution\
to the general Riemann-Hilbert problem discussed in \KSW.
Fortunately, the case of movable branch points \solgen\ is much simpler.
Introducing a continuous coordinate
$h={1\over N}h_i$ and a density $\rho(h)=\partial i/\partial h_i$,
one finds, with the help of \xidef,
\eqn\saddle{\barint_b^a dh' {\rho(h') \over h-h'} = \log(h-b) + {1 \over 2} A_T +\half \tilde{V}'(h),}
where the effective potential $\tilde{V}'(h)$ is, for
a finite number of non-zero $\tau_n$'s, a polynomial in $h$
whose coefficients are selfconsistently dependent on the
first few moments $H_n=N^{-1 -n} \sum h_i^n$  of the resolvent
$H(h)=\int_0^a dh' {\rho(h') \over h-h'}$. 
It is found as follows. Define the potential
\eqn\pot{
V(h)=\sum_{k=2}^{\infty} {1 \over k} \tau_k h^k.}
Then
\eqn\effpot{
\tilde{V}'(h)= {1 \over 2 \pi i} \oint {ds  \over (h - s)^2}~
~V(s~e^{- H(s)}).}
Here the contour surrounds the cut $[b,a]$ of $H(h)$.
The simplest non-trivial example consists of taking
$\tau_k=0$ for $k\geq3$. The analysis then becomes very
similar to the one for chiral $YM_2$, obtained in \WC.
Eq.\saddle\ then reads:
\eqn\simpsd{
\barint_b^a\ {dh'\rho(h')\over h-h'}=
           \ln (h-b)\ +\ {A_T\over 2}+{\t_2\over 2}(1-h).
}
The term $\log(h-b)$ is a 
consequence\foot{A brief argument shows that for
the phase corresponding to the sum over surfaces, a part of the
density, starting at the origin and finishing at a point $b$, is
saturated at it's maximum value. The sum over coverings is an
expansion in powers of $e^{-A_T}$ and $\tau_2$. For small $\tau_2$ and
small $e^{-A_T}$ i.e. large $A_T$ the $e^{-A_T\sum_i h_i}$ term in
\solgen\ attracts all the $h_i$ towards the origin saturating the
constraint $h_{i+1}<h_i$, $\rho(h)\leq 1$.} 
of the fact that the $h_i$ are a set of ordered integers.
The density is therefore constrained to have a maximum value of 1 (see \KSW).
This equation is solved in the standard way by a contour integral
\eqn\contH{\eqalign{
H(h)&=\int_0^a \ {dh'\rho(h')\over h-h'}\cr
    &=\ln\big({h\over h-b}\big)+
      \sqrt{(h-a)(h-b)}\oint \ {ds\over 2\pi i}
{\ln(s-b)+{A_T\over 2}+{\t_2\over 2}(1-h)\over
    (h-s)\sqrt{(s-a)(s-b)}}.\cr
}}
Expanding the contour out to infinity we catch a pole at $s=h$, the
discontinuity across the cut of the logarithm, and a contribution from the
contour at infinity. The final result is
\eqn\Hofh{\eqalign{
H(h)=&\ln h\ +\ {A_T\over 2}-{\t_2\over 2}+
{\t_2\over 2} 
(h-\sqrt{(h-a)(h-b)})\cr
&-2\ln\bigg[{\sqrt{h-a}+\sqrt{h-b}\over\sqrt{a-b}}\bigg].}
}
The density $\rho(h)$ is as given by the discontinuity of $H(h)$ across
its cut: \foot{Specifically
$\rho(h)={i\over 2\pi}(H(h+i\epsilon)-H(h-i\epsilon))$ where
$\epsilon$ is a small positive real number.}
\eqn\dens{
\rho(h)={2\over \pi}\cos^{-1}\bigg(\sqrt{h-b\over a-b}\bigg)
        -{\t_2\over 2}\sqrt{(h-a)(h-b)},
}
and the cut points $a$ and $b$ are determined from the behaviour of
$H(h)$ for large $h$:
\eqn\asympH{
H(h)={1\over h}+{\cal O}\big({1\over h^2}\big).
}
Imposing this asymptotic behaviour on \Hofh\ 
leads to the two equations
\eqn\bcs{
\chi=\t_2^2 e^{-A_T}~e^{\chi} \ \ \ {\rm and} \ \ \ \eta-\t_2 
e^{-A_T+\chi}=1.
}
where we have defined $\chi=A_T+2\log((a-b)/4)$ and $\eta=(a+b)/2$. 
Differentiating \solgen\ w.r.t. to $A_T$ leads to
${\partial\over\partial A}{\cal F}=-<h>+{1\over 2}$ where the free
energy ${\cal F}$ is defined here by ${\cal F}=1/N^2\ln{\cal Z}$. The
expectation value $<h>=\int_0^adh\rho(h)h$ can be calculated from the
expansion for $H(h)$ \Hofh . Using \bcs\ it is possible to integrate
up to obtain an explicit expression for the spherical contribution to
${\cal F}$
\eqn\free{
{\cal F}=e^{-A_T+\chi}\big(1-{3\over 4}\chi
                                        +{1\over 6}\chi^2\big).
}
>From \bcs\ it is clear that $\chi$ is a power series in $\t^2_2 e^{-A_T}$.  
We can thus perform a standard Lagrange inversion on \bcs ,\free,
and obtain after a short calculation:
\eqn\freesol{
{\cal F}=\sum_{n=1}^{\infty}
{n^{n-3} \over n!}~\t_2^{2n-2}~e^{-n A_T}.}
This result was first obtained in \Wati,\WC. 
To discuss the convergence properties of \freesol,
it is natural to take the
continuum coupling proportional to $A_T$, since the branch
points can be located anywhere on the manifold (see also \tautwo):
$\t_2=t A_T$. Note that the series is only
convergent for $t^2 A_T^2 e^{-A_T}<e^{-1}$. Beyond this point the boundary
conditions \bcs\ lead to a non-physical complex value for 
$\chi$. We see that the sum over
branched coverings is convergent for both large and small areas. 
For large enough $t$ and intermediate values of the area $A_T$,
however, the entropy of the branch points is sufficient to cause the
sum to  diverge. 
It is interesting to understand this divergence in terms of the Young
tableau density $\rho(h)$. 
Along the critical line $(\t_2,A_T)$ and $\t_2>0$  
the density becomes flat at its upper end
point $a$, i.e. the singularity at the end point changes from
$\rho(h)\sim (a-h)^{1/2}$ to $\rho(h)\sim (a-h)^{3/2}$. 
Along the critical line $(\t_2,A_T)$ and $\t_2<0$
it is the singularity of the density at the point $b$ that
changes from $1/2$ to $3/2$. 
This is just as occurs in matrix models of
pure $2D$ gravity, and indeed for $t^2 A^2 e^{-A_T}\sim e^{-1}$ the free
energy \freesol\ behaves as 
\eqn\sterling{
{\cal F} \sim(e^{-1} - t^2 A_T^2 e^{-A_T})^{5/2}.}

So far in our analysis we have ignored the constraint that we are
summing over positive Young tableaux i.e. that $b>0$. Since $b=0$ is
not a singular point in the boundary conditions \bcs\ it does not
correspond to a singularity in the sum over surfaces. If, however, we
take the sum over representations \solgen\ as fundamental (as would be
the case for QCD$_2$) then one should take this constraint into
account.

\vskip 50pt
\hskip 55pt 
\epsfbox{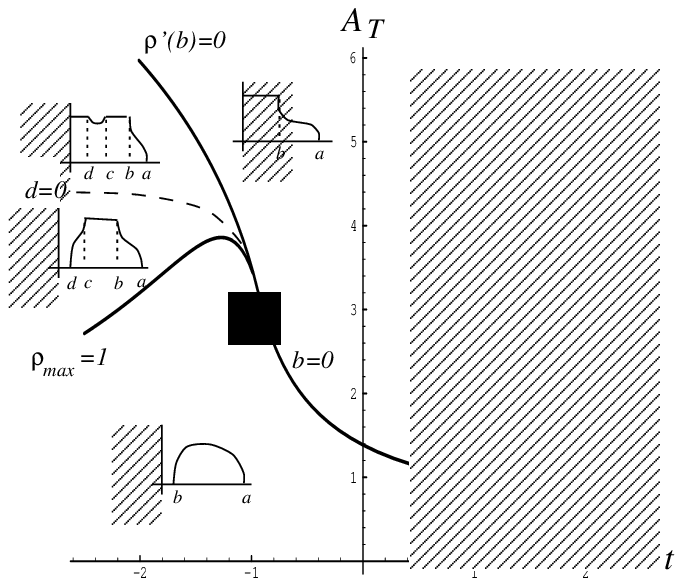}
\vskip -1pt
\vskip 10pt
\centerline{
Fig 2. Phase diagram in the $t, A_T$ plane.}
\centerline{A typical density, $\rho(h)$, is sketched in each phase.} 
\vskip 20pt

Setting $b=0$ in \bcs\ leads to the pair of equations
\eqn\beqo{
t^2 A^2 e^{-A_T}=\chi e^{-\chi} \ \ \ {\rm and} \ \ \ A_T=\chi+2\ln(2-\sqrt{\chi}).
}
which determine $t$ and $A_T$
parametrically in terms of $\chi$. This curve is plotted in
Fig. 2. 
It connects the dot on the left to the shaded area on the
right. Immediately below this line the support of the density is
entirely positive, i.e. it starts at a positive value of $h$. In
addition it is less than one for the full range of its support, see
Fig. 2. The phase transition across this line is exactly analogous
to the large $N$ phase transition that occurs in QCD (see \GW\GB
). There there is a phase transition separating a strong coupling
regime from a weak coupling phase, and as is the case here, there is
no singularity in either phase indicating the the transition
point. Indeed it is possible to formulate the Gross-Witten \GW\ and
Br\'ezin-Gross \GB\ models in terms of a sum over representations \MW\
and the large $N$ phase transition is precisely the point at which the
density just begins to touch (or pull away from) the origin.
Further analysis shows that there are two further phases for the
model, as first observed in \WC.
There is one phase where the density has an entirely positive support
but attains its maximal value $\rho(h)=1$ over a single finite
interval, and another phase where the density starts at the origin and
has two separated intervals where $\rho(h)=1$. Both of these phases
involve two separated nontrivial cuts and can be calculated in terms
of elliptic functions. We do not present here explicit results for any
of these extra phases. The complete phase diagram for the partition
function is shown in Fig. 2. In each phase is sketched a
typical density. Along the transition lines are indicated the
corresponding critical behaviours of the density. The exact position
of the line along which the point $d$ equals zero has not been
calculated. We indicate this by the using a dotted line. The sum over
representations is thus seen to have a much richer phase structure than the perturbative sum over surfaces it generates.

\newsec{Concluding remarks}

We have introduced a new class of matrix gauge models which
solve the general problem of counting branched covers of
orientable two-dimensional manifolds with specified branch point 
structure. The result is given as a weighted sum over
polynomial representations of $U(N)$. We conjecture that
a similar methodology could be applied to non-orientable
surfaces by replacing the complex matrices by real matrices
and the group $U(N)$ by $O(N)$. 

Our approach actually treats also  the 
case of  manifolds with boundaries.   Indeed, due to the invariance with respect to 
area-preserving diffeomorphisms, the problem does not depend on the length of the 
boundaries and the latter can be considered as punctures.

 As an interesting by-product of our investigation,
we have derived in a constructive and transparent
fashion important features of
the original Gross-Taylor interpretation of $YM_2$ as a string theory.

An interesting problem which we did not consider in this paper is the calculation of Wilson loops.
The Wilson loop functional defined as the sum of all nonfolding branched surfaces spanning a 
  closed contour $\CC\in \MG$ should satisfy a set of loop equations with a contact term similar
to those considered in \ref\KK{V. Kazakov and I. Kostov, \np B 176 (1980) 199.}.

\vfil\eject


\appendix{A}{Free energies for $\CN0$ fixed, identical branch points}

In this appendix we give some examples of how to extract the 
combinatorics of connected coverings of a smooth manifold
with $\CN0$ fixed, identical branch points. The partition sum \solution\ 
for this case becomes
\eqn\lattice{
Z=\sumh 
\ \bigg({\Delta_h \over \Oh}\bigg)^{2-2 G}~     
\bigg[  {\char(B) \over \char(A_1)}
\bigg]^{\CN0}~e^{-|h| A_T}. }
Define
\eqn\zndef{Z=1+\sum_{n=1}^{\infty}Z_n~e^{-n A_T}.}
Using the Frobenius formula for characters,
the first few $Z_n$'s, up to order four, are found to be:
\eqn\zn{\eqalign{
Z_1=N^{2-2 G}~t_1^{\CN0}~~~&  \cr
Z_2=(\half N^2)^{2-2 G}~\big[&(t_1^2+{1 \over N}t_2)^{\CN0}+
(t_1^2-{1 \over N}t_2)^{\CN0}\big] \cr
Z_3=({1 \over 6}N^3)^{2-2 G}~\big[&(t_1^3+{3 \over N} t_1 t_2 +
{2 \over N^2} t_3)^{\CN0}+
(t_1^3-{3 \over N} t_1 t_2 + {2 \over N^2} t_3)^{\CN0}\big] +\cr
  + ({1 \over 3}N^3)^{2-2 G}~\big[&(t_1^3-{1 \over N^2} t_3)^{\CN0}\big] \cr
Z_4=({1 \over 24} N^4)^{2-2 G}~\big[&(t_1^4+{6 \over N}t_1^2 t_2+
{8 \over N^2} t_1 t_3 + {3 \over N^2} t_2^2+ {6 \over N^3} t_4)^{\CN0}+\cr
&~ +(t_1^4-{6 \over N}t_1^2 t_2+
{8 \over N^2} t_1 t_3 + {3 \over N^2} t_2^2- {6 \over N^3} t_4)^{\CN0}\big]+ \cr
+ ({1 \over 8}~N^4)^{2-2 G}~\big[&(t_1^4+{2 \over N}t_1^2 t_2-
{1 \over N^2}t_2^2-{2\over N^3}t_4)^{\CN0}+
(t_1^4-{2 \over N}t_1^2 t_2-
{1 \over N^2}t_2^2
+{2\over N^3}t_4)^{\CN0}\big] +\cr
+ ({1 \over 12}N^4)^{2-2 G}~\big[&(t_1^4-{4 \over N^2} t_1 t_3 +
{3 \over N^2}t_2^2)^{\CN0}\big] .}}
These expressions allow us to obtain explicit
results for the map counting problem. The free energy counting
connected surfaces is
\eqn\fndef{F=\log Z=\sum_{n=1}^{\infty} F_n~e^{-n A_T}.} 
Let us present some
explicit low order results:

$G=0$:
\eqn\fnsphere{\eqalign{
F_1=& N^2~t_1^{\CN0} \cr
F_2=& \sum_{g=0}^{[\half \CN0]-1} N^{2-2 g} 
\half \Big({\CN0 \atop 2 g+2}\Big) (t_1^2)^{\CN0-2 g-2}
t_2^{2 g+2} \cr
F_3=& N^2 \Bigg[ 4 \Big({\CN0 \atop 4}\Big) t_1^{3 \CN0-8} t_2^4+
\Big({\CN0 \atop 2,1}\Big) t_1^{3 \CN0-7} t_2^2 t_3  +
{1 \over 3} \Big({\CN0 \atop 2}\Big) t_1^{3 \CN0-6} t_3^2 \Bigg]+ \cr
&~+N^0 \Bigg[ 40 \Big({\CN0 \atop 6}\Big) t_1^{3 \CN0 -12} t_2^6+ 
{3 \over 2} \Big({\CN0 \atop 2,2,1}\Big) t_1^{3 \CN0-11} t_2^4 t_3  + \cr
&~~~\;\;\;\;\;\;\;+2 \Big({\CN0 \atop 2,2}\Big) t_1^{3 \CN0-10} t_2^2 t_3^2 +
{1 \over 3} \Big({\CN0 \atop 3}\Big) t_1^{3 \CN0-9} t_3^3 \Bigg]+ 
{\cal O}(N^{-2}) \cr
F_4=& N^2 \Bigg[ \Bigg( 124 \Big({\CN0 \atop 6}\Big) 
+ 13 \Big({\CN0 \atop 1,4}\Big) 
+{5 \over 4} \Big({\CN0 \atop 2,2}\Big) +
{5 \over 16} \Big({\CN0 \atop 3}\Big) -
{1 \over 16} \Big({2 \CN0 \atop 6}\Big)
\Bigg) t_1^{4 \CN0-12} t_2^6 + \cr
&~~~+\Bigg( 27 \Big({\CN0 \atop 1,4}\Big) +
3 \Big({\CN0 \atop 1,1,2}\Big) \Bigg) t_1^{4 \CN0-11} t_2^4 t_3 +\cr
&~~~+\Bigg( 6 \Big({\CN0 \atop 2,2}\Big) +
\Big({\CN0 \atop 2,1}\Big) \Bigg) t_1^{4 \CN0-10} t_2^2 t_3^2 +\cr
&~~~+ \Big({\CN0 \atop 1,1,1}\Big) t_1^{4 \CN0-9} t_2 t_3 t_4 +
\Big({\CN0 \atop 3}\Big) t_1^{4 \CN0-9} t_3^3+
{1 \over 4} \Big({\CN0 \atop 2}\Big) t_1^{4 \CN0-8} t_4^2+ \cr
&~~~+ \Bigg( 4 \Big({\CN0 \atop 3,1}\Big) +
\half  \Big({\CN0 \atop 1,1,1}\Big) \Bigg) t_1^{4 \CN0-10} t_2^3 t_4
\Bigg] + {\cal O}(N^0).}}

$G=1$:
\eqn\fntorus{\eqalign{
F_1=& t_1^{\CN0} \cr
F_2=& {3 \over 2} t_1^{2 \CN0}+
\sum_{g=2}^{[\half \CN0]+1} N^{2-2 g} 2
\Big({\CN0 \atop 2 g-2 }\Big) (t_1^2)^{\CN0 +2 -2 g} t_2^{2 g-2} \cr
F_3=&{4 \over 3} t_1^{3 \CN0} + N^{-2} \Bigg[
16 \Big({\CN0 \atop 2  }\Big) t_1^{3 \CN0 -4} t_2^2 +
3 \CN0 t_1^{3 \CN0 -3} t_3 \Bigg] + {\cal O}(N^{-4}) \cr
F_4=&{7 \over 4} t_1^{4 \CN0} + N^{-2} \Bigg[
\Bigg(7 \CN0 + 60 \Big({\CN0 \atop 2  }\Big) \Bigg)
t_1^{4 \CN0 -4} t_2^2 +
9 \CN0 t_1^{4 \CN0 -3} t_3 \Bigg] + {\cal O}(N^{-4}).}}

$G=2$:
\eqn\fntwotor{\eqalign{
F_1=& N^{-2} t_1^{\CN0} \cr
F_2=& N^{-4} {15 \over 2} t_1^{2 \CN0}+
\sum_{g=4}^{[\half \CN0]+3} N^{2-2 g} 8
\Big({\CN0 \atop 2 g-6 }\Big) (t_1^2)^{\CN0 +6 -2 g} t_2^{2 g-6} \cr
F_3=&N^{-6} {220 \over 3} t_1^{3 \CN0} + N^{-8} \Bigg[
640 \Big({\CN0 \atop 2  }\Big) t_1^{3 \CN0 -4} t_2^2 +
135 \CN0 t_1^{3 \CN0 -3} t_3 \Bigg] + {\cal O}(N^{-10}) \cr
F_4=&N^{-8} {5275 \over 4} t_1^{4 \CN0} + N^{-10} \Bigg[
\Bigg(3760 \CN0 + 41280 \Big({\CN0 \atop 2  }\Big) \Bigg)
t_1^{4 \CN0 -4} t_2^2 +
8505 \CN0 t_1^{4 \CN0 -3} t_3 \Bigg] + \cr
&~+{\cal O}(N^{-12}).}}
Here have employed the standard notation for
binomial and multinomial coefficients.
It is instructive to draw 
the Riemann surfaces corresponding to the 
various terms in \fnsphere,\fntorus,\fntwotor. As will be seen, the 
combinatorics involved increases rapidly in complexity.

The above examples are easily checked to be in agreement
with the Riemann-Hurwitz formula. Of course, this formula
only gives a {\it necessary} condition for the existence
of a Riemann surface.
The above results can be used to decide whether a Riemann surface
of given $G$,$g$ and branch point structure 
$t_2^{{\cal N}_2} t_3^{{\cal N}_3} \ldots$ actually {\it exists}.
For example, we see from $F_4$ in \fnsphere\ that there
exists a fourfold cover of the sphere by a sphere with exactly six
simple branch points, where two branchpoints are located at
each of three locations of the target manifold (the associated
symmetry factor is ${5 \over 16}$).
As a second example, we see from $F_3$ in \fntorus\ that 
there exists a triple cover of the torus by a double torus
with exactly one branch point of order 2 (the associated
symmetry factor is 3).

\appendix{B}{Free energies for an arbitrary number of movable branch points}

In this appendix we give some concrete examples of how the maps are
counted in the continuum limit we defined above.
Define
\eqn\zndef{\CZ=1+\sum_{n=1}^{\infty}\CZ_n~e^{-n A_T}.}
The first few auxiliary $\xi^h_k$ (see \xidef) are:
\eqn\xifew{\eqalign{
\xi^h_{1} & =\sum_i h_i - {1 \over 2} N (N-1) \cr
\xi^h_{2} & ={1 \over 2} \sum_i h_i^2 - {2 N-1 \over 2} \sum_i h_i+
{N(N-1)(2 N-1) \over 6} \cr
\xi^h_{3} & ={1 \over 3} \sum_i h_i^3 - {1 \over 2} \sum_i h_i \sum_j h_j-
{2 N -1 \over 2} \sum_i h_i^2 + \cr
& + {1 \over 3} ({9 \over 2} N^2 - {9 \over 2} N +2) \sum_i h_i
- {N(N-1)(3 N^2 - 3 N + 2) \over 8} }}
They are easily computed from the following representation of 
eq.\xidef:
\eqn\xicont{
\xi_n^h={1 \over n^2} \oint {dh \over 2 \pi i}
~h(h-1) \ldots (h-n+1)~\exp\big[
\sum_{p=1}^{\infty} {(-n)^p \over p!} 
\big({\partial \over \partial h}\big)^{p-1}
H(h)\big],}
where we have introduced $H(h)=\sum_{i=1}^{N} {1 \over h-h_i}$.
The first few $\CZ_n$'s, up to order four, are:
\eqn\zn{\eqalign{
\CZ_1=&N^{2-2 G}  \cr
\CZ_2=&(\half N^2)^{2-2 G}~\big(
e^{{1 \over N} \tau_2} + e^{-{1 \over N} \tau_2} \big) \cr
\CZ_3=&({1 \over 6}N^3)^{2-2 G}~\big(
e^{{3 \over N} \tau_2 + {2 \over N^2} \tau_3} +
e^{-{3 \over N} \tau_2 + {2 \over N^2} \tau_3} \big) +\cr
  &~+ ({1 \over 3}N^3)^{2-2 G}~e^{-{1 \over N^2} \tau_3} \cr
\CZ_4=&({1 \over 24} N^4)^{2-2 G}~\big( 
e^{{6 \over N} \tau_2 + {8 \over N^2} \tau_3 + {6 \over N^3} \tau_4}
+e^{-{6 \over N} \tau_2 + {8 \over N^2} \tau_3  -{6 \over N^3} \tau_4}\big)
+ \cr
&~+ ({1 \over 8}~N^4)^{2-2 G}~\big(
e^{{2 \over N} \tau_2 - {2 \over N^3} \tau_4}+
e^{-{2 \over N} \tau_2 + {2 \over N^3} \tau_4}\big) +\cr
&+ ({1 \over 12}N^4)^{2-2 G}~e^{-{4 \over N^2} \tau_3} .}}
This gives the following continuum free energies (as can be checked
easily from the results of appendix A):

$G=0$:
\eqn\cfnsphere{\eqalign{
\CF_1=& N^2 \cr
\CF_2=& \sum_{g=0}^{\infty} N^{2-2 g} 
\half {1 \over (2 g+2)!}
\t_2^{2 g+2} \cr
\CF_3=& N^2 \Big[ {1 \over 6} \t_2^4+
\half \t_2^2 \t_3 + {1 \over 6}  \t_3^2 \Big] +\cr
&~+ \Big[ {1 \over 18} \t_2^6 +
{3 \over 8} \t_2^4 \t_3 + {1 \over 2}  \t_2^2 \t_3^2 
+ {1 \over 18}  \t_3^3 \Big] +
{\cal O}(N^0) \cr
\CF_4=& N^2 \Big[ {1 \over 6} \t_2^6 +
{27 \over 24} \t_2^4 \t_3 +
{3 \over 2} \t_2^2 \t_3^2 + \t_2 \t_3 \t_4 +
{1 \over 6} \t_3^3+
+ {2 \over 3} \t_2^3 \t_4 +
{1 \over 8}  \t_4^2
\Big] + {\cal O}(N^0).}}

$G=1$:
\eqn\cfntorus{\eqalign{
\CF_1=& 1 \cr
\CF_2=& {3 \over 2} +
\sum_{g=2}^{\infty} N^{2-2 g} {2 \over (2 g-2)!}
\t_2^{2 g-2} \cr
\CF_3=&{4 \over 3} + N^{-2} \Big[
8 \t_2^2 +
3 \t_3 \Big] + {\cal O}(N^{-4}) \cr
\CF_4=&{7 \over 4} + N^{-2} \Big[
30 \t_2^2 + 9 \t_3 \Big] + {\cal O}(N^{-4}).
}}

$G=2$:
\eqn\cfntwotor{\eqalign{
\CF_1=& N^{-2} \cr
\CF_2=& N^{-4} {15 \over 2} +
\sum_{g=4}^{\infty} N^{2-2 g} {8 \over (2 g-6)!}
\t_2^{2 g-6} \cr
\CF_3=& N^{-6} {220 \over 3} + N^{-8} \Big[
320 \t_2^2 +
135 \t_3 \Big] + {\cal O}(N^{-10}) \cr
\CF_4=& N^{-8} {5275 \over 4} + N^{-10} \Big[
20640 \t_2^2 + 8505 \t_3 \Big] + {\cal O}(N^{-12}).
}} 

\listrefs

\bye